\documentclass{article}
\usepackage{spconf,amsmath,graphicx}
\usepackage{epstopdf}
\usepackage{epsfig}
\usepackage{amsfonts}
\usepackage{amssymb}
\newcommand{\argmin}{\operatornamewithlimits{argmin}}
\usepackage{amsthm}

\usepackage{amsmath}
\usepackage{amsfonts}
\usepackage{amssymb}
\usepackage{latexsym}
\usepackage{caption}
\usepackage{subcaption}
\usepackage{algorithm}
\usepackage{hyperref}
\usepackage[noend]{algpseudocode}
\usepackage{color}
\makeatletter
\def\BState{\State\hskip-\ALG@thistlm}
\makeatother

\usepackage[left]{lineno}
\title{Mahalanobis Distance for Class Averaging of Cryo-EM Images}
 \name{Tejal Bhamre$^1$ \qquad\qquad\qquad\qquad Zhizhen Zhao$^2$ \qquad\qquad\qquad Amit Singer$^3$\sthanks{This work was partially supported by Award Number
R01GM090200 from the NIGMS, FA9550-12-1-0317 from AFOSR, Simons Foundation Investigator Award and Simons Collaborations on Algorithms and Geometry, and the Moore Foundation Data-Driven Discovery Investigator Award.}}
\address{\small{$^1$Dept. of Physics, Princeton University; $^2$ Dept. of  Electrical Engineering, UIUC; $^3$Dept. of Mathematics, Princeton University}}

\begin{document}
%\ninept
%
\maketitle
\begin{abstract}
Single particle reconstruction (SPR) from cryo-electron microscopy (EM) is a technique in which the 3D structure of a molecule needs to be determined from its contrast transfer function (CTF) affected, noisy 2D projection images taken at unknown viewing directions. One of the main challenges in cryo-EM is the typically low signal to noise ratio (SNR) of the acquired images. 2D classification of images, followed by class averaging, improves the SNR of the resulting averages, and is used for selecting particles from micrographs and for inspecting the particle images. We introduce a new affinity measure, akin to the Mahalanobis distance, to compare cryo-EM images belonging to different defocus groups. The new similarity measure is employed to detect similar images, thereby leading to an improved algorithm for class averaging. We evaluate the performance of the proposed class averaging procedure on synthetic datasets, obtaining state of the art classification. 
\end{abstract}

\section{Introduction}
\label{sec:intro}
\begin{figure}
\begin{center}
\vspace{-.2in}
\includegraphics[width=\columnwidth]{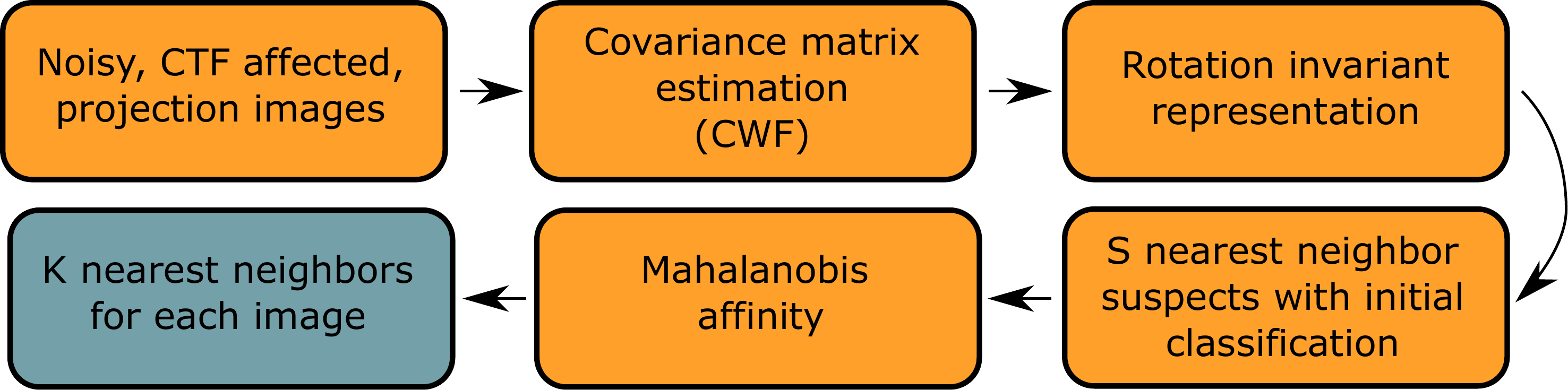}
\vspace{-.2in}
\caption{Pipeline of algorithm}\label{fig:pipeline}
\end{center}
\vspace{-.2in}
\end{figure}
 
SPR from cryo-EM is a rapidly advancing technique in structural biology to determine
the 3D structures of macromolecular complexes in their native state \cite{Frank1, Kuhlbrandt1443},
 without the need for crystallization. First, the sample, consisting of randomly oriented, nearly identical copies of a macromolecule, is frozen in a thin ice layer. An electron microscope is used to acquire top view images of the sample, in the form of a large image called a `micrograph', from which individual particle images are picked semi-automatically. After preprocessing the selected raw particle images, the images are clustered. The images within each class are averaged to enjoy a higher SNR than the individual images. This step is known as ``class averaging". To minimize radiation damage, cryo-EM imaging must be constrained to low electron doses, which results in a very low SNR in the acquired 2D projection images. Class averaging is thus a crucial step in the SPR pipeline: class averages are used for a preliminary inspection of the dataset, to eliminate outliers, and in semi-automated particle picking \cite{relion}. Typically, a user manually picks particles from a small number of micrographs. These are used to compute class averages, which are further used as templates to pick particles from all micrographs. Class averages are also used in subsequent stages of the SPR pipeline, such as 3D ab-initio modeling.
 
The two popular approaches for 2D class averaging \cite{Penczek1992,Penczek1996, vanHeel1990a, vanHeel1981, chirikjian1, chirikjian2} in cryo-EM are multivariate statistical analysis (MSA)\cite{vanHeel1981} with multi-reference alignment (MRA) \cite{Dube1993} and iterative reference-free alignment using K-means clustering \cite{Penczek1996}. Popular cryo-EM packages like RELION, XMIPP, EMAN2, SPIDER, SPARX, IMAGIC \cite{imagic, spider, eman2, xmipp, relion, gpurelion} use some of these methods for class averaging. RELION uses a maximum likelihood classification procedure. A faster and more accurate approach for 2D class averaging based on rotationally invariant representation was introduced in \cite{zhao} and is implemented in the cryo-EM software package ASPIRE (\url{http://spr.math.princeton.edu/}).

Recently in \cite{cwf}, it was shown that preliminary inspection of the underlying clean images and outlier detection can be performed at an earlier stage, by better denoising the acquired images using an algorithm called Covariance Wiener Filtering (CWF). In CWF, the covariance matrix of the underlying clean projection images is estimated from their noisy, CTF-affected measurements. The covariance is then used in the classical Wiener deconvolution framework to obtain denoised images. 

There are two main contributions of this paper. First, we introduce a new similarity measure, related to the Mahalanobis distance \cite{mah}, to compute the similarity of pairs of cryo-EM images. Second, we use the proposed Mahalanobis distance to improve the class averaging algorithm described in \cite{zhao}. We first obtain for each image a list of $S$ other images suspected as nearest neighbors using the rotation invariant representation (see Sec.~\ref{sec:section2} for details), and then rank these suspects using the Mahalanobis distance (see Fig.~\ref{fig:pipeline}). The top $K$ nearest neighbors, where $K<S$, given by this procedure are finally aligned and averaged to produce class averages. We test the new algorithm on a synthetic dataset at various noise levels and observe an improvement in the number of nearest neighbors correctly detected.

\begin{figure}
\begin{center}
\vspace{-.2in}
\includegraphics[scale=0.35]{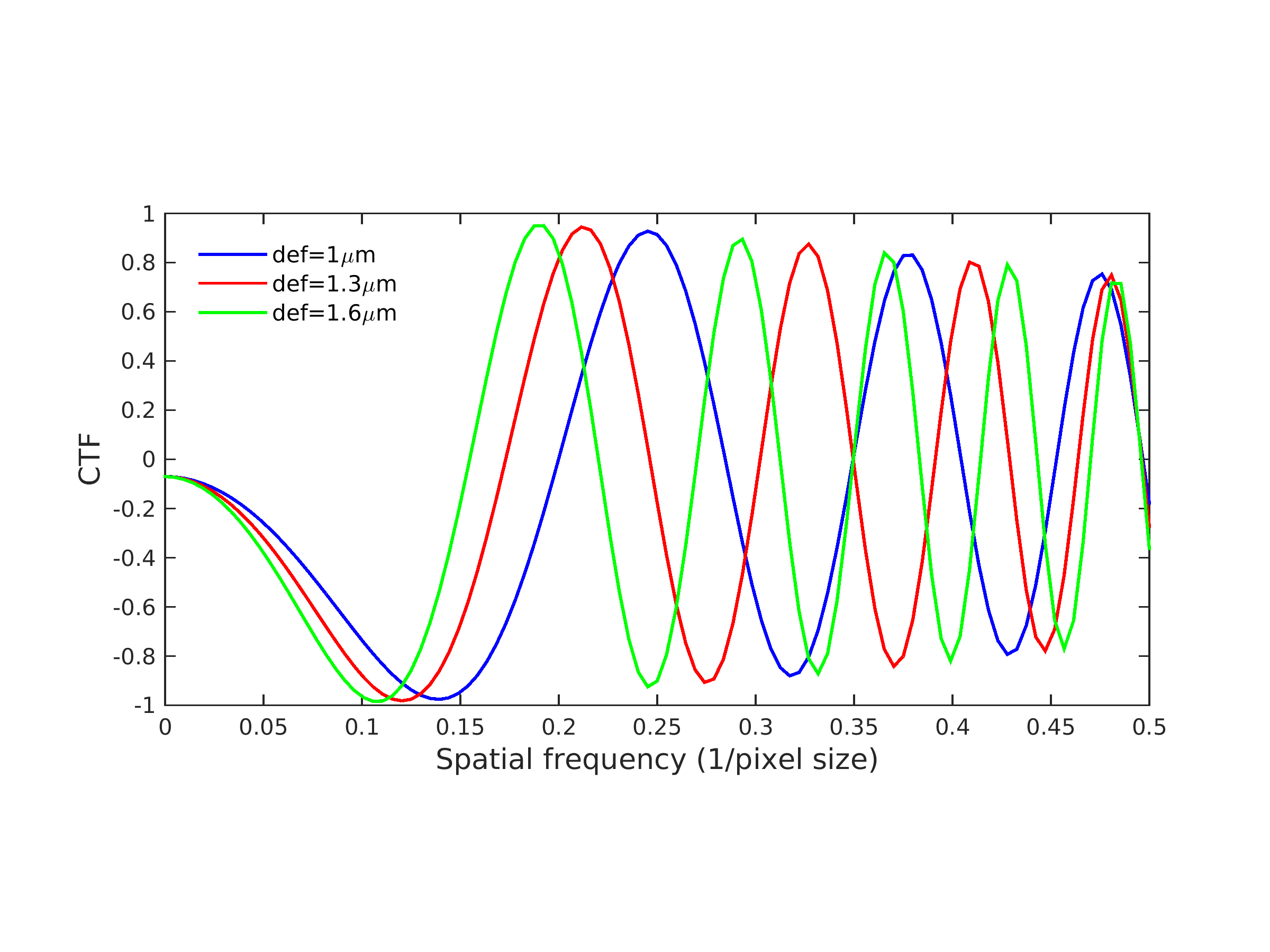}
\vspace{-.5in}
\caption{CTF's for different values of the defocus. CTF parameters used are: the amplitude contrast $\alpha = 0.07$
, the electron wavelength $\lambda= 2.51pm$, the spherical aberration constant $Cs=2.0$
, the B-factor $B=10$, the defocus$=1\mu m, 1.3\mu m,$ and
$1.6\mu m$, and the pixel size is $2.82\AA$. See 
eq.~\eqref{eq:ctf}}\label{fig:ctf}
\end{center}
\vspace{-.2in}
\end{figure}

\section{Background}
\label{sec:section2}
\subsection{Image Formation Model}
Under the linear, weak phase approximation (see \cite[Chapter~2]{frankbook}), the image formation model in cryo-EM is given by
\begin{equation}
y_i = a_i\star x_i + n_i
\label{eq:imreal}
\end{equation}
where $\star$ denotes the convolution operation, $y_i$ is the noisy projection image in real space, $x_i$ is the underlying clean projection image in real space, $a_i$ is the point spread function of the microscope, and $n_i$ is additive Gaussian noise that corrupts the image. In the Fourier domain, images are multiplied with the Fourier transform of the point spread function, called the CTF, and eqn.(\ref{eq:imreal}) can be rewritten as
\begin{equation}
Y_i = A_iX_i + N_i
\label{eq:imfour}
\end{equation}
where $Y_i$, $X_i$ and $N_i$ are the Fourier transforms of $y_i$, $x_i$ and $n_i$ respectively.
The CTF is approximately  given by (see  \cite[Chapter~3]{frankbook})
\begin{equation}
CTF(\hat{k};\Delta\hat{z}^2)= e^{-{B\hat{k}}^2}\sin[-\pi \Delta\hat{z}\hat{k}^2 + \frac{\pi}{2} \hat{k}^4]
\label{eq:ctf}
\end{equation}
where 
$\Delta\hat{z}=\frac{\Delta z}{[C_s \lambda]^{\frac{1}{2}}}$ is the ``generalized defocus" and $\hat{k}=[C_s \lambda]^{\frac{1}{4}}k$ is the ``generalized spatial frequency", and $B$ is the B-factor for the Gaussian envelope function. CTF's corresponding to different defocus values have different zero crossings (see Fig.\ref{fig:ctf}). Note that the CTF inverts the sign of the image's Fourier coefficients when it is negative, and completely suppresses information at its zero crossings.

\subsection{Rotationally Invariant Class Averaging}
The procedure for class averaging, described in \cite{zhao}, was demonstrated to be both faster and more accurate than other existing class averaging procedures. It consists of three main steps. First, principal component analysis (PCA) of CTF-corrected phase flipped images is computed. We refer to this step as steerable PCA, because the procedure takes into account that the 2D covariance matrix commutes with in-plane rotations. Second, the bispectrum of the expansion coefficients in the reduced steerable basis is computed. The bispectrum is a rotationally invariant representation of images, but is typically of very high dimensionality. It is projected onto a lower dimensional subspace using a fast, randomized PCA algorithm \cite{rokhlin}. One way to compare images after this step is using the normalized cross correlation. At low SNR, it is difficult to identify true nearest neighbors from the cross correlation. Therefore, Vector Diffusion Maps (VDM) [15] was used to further improve the initial classification by taking into account the consistency of alignment transformations among nearest neighbor suspects. 
\vspace*{-\baselineskip}

\subsection{Covariance Wiener Filtering (CWF)}
CWF was proposed in \cite{cwf} as an algorithm to (i) estimate the CTF-corrected covariance matrix of the underlying clean images (since phase flipping is not an optimal correction)  and (ii) using the estimated covariance to solve the associated deconvolution problem in eqn. \ref{eq:imfour} to obtain denoised images, that are estimates of $X_i$ for each $i$ in 
eqn. \ref{eq:imfour}. The first step involves estimating the mean image of the dataset, $\mu$, denoted $\hat{\mu}$, followed by solving a least squares problem to estimate the covariance $\Sigma$, denoted $\hat{\Sigma}$. Under the assumption of additive white Gaussian noise with variance $\sigma^2$, the estimate of the underlying clean image $X_i$ is given by
\begin{equation}
\hat{X_i}=(I-H_iA_i)\hat{\mu} + H_iY_i
\end{equation}
where $H_i=\hat{\Sigma}A_i^T(A_i \hat{\Sigma}A_i^T + \sigma^2 I)^{-1}$
%
%\subsection{Fourier Bessel Steerable PCA}
%For both class averaging and CWF, the images are expressed in the Fourier Bessel basis. Since the basis elements are outer products of radial functions and angular Fourier modes, the covariance matrix is block diagonal in this choice of basis. Therefore each block of the covariance, corresponding to a particular angular frequency, can be computed separately, making the computation more efficient.
\section{Anisotropic Affinity}
The Mahalanobis distance in statistics \cite{mah} is a generalized, unitless and scale invariant similarity measure that takes correlations in the dataset into account. It is popularly used for anomaly detection and clustering \cite{mahclust1, mahclust2}.
 
Our goal is to define a similarity measure to compare how close any two cryo-EM images are, given the CTF-affected, noisy observations for a pair of images from possibly different defocus groups, say $Y_i$ and $Y_j$ in eq. \ref{eq:imfour}.
CTF correction is a challenging problem due to the numerous zero crossings of the CTF. A popular, albeit, only partial correction of CTF is phase flipping, which involves simply inverting the sign of the Fourier coefficients. This corrects for the phase inversion caused by the CTF, but does not perform amplitude correction. Since phase flipping is suboptimal as a method for CTF correction, computing nearest neighbors using the Euclidean distance between features constructed from phase flipped, denoised images can suffer from incorrectly identified neighbors. One simple approach would be to use the Euclidean distance between the CWF denoised images, as a measure of similarity. However, the optimality criterion for obtaining CWF denoised images is different from that for identifying nearest neighbors. Also, after CWF denoising, noise is no longer white, so the Euclidean distance is not an optimal measure of affinity.  

In our statistical model, the underlying clean images $X_1, X_2, \ldots X_n \in \mathbb{C}^{d}$ (where $n$ is the total number of images and $d$ is the total number of pixels in each image) are assumed to be independent, identically distributed (i.i.d.) samples drawn from a Gaussian distribution. Further, we assume that the noise in our model is additive white Gaussian noise 
\begin{eqnarray} 
X_i  \sim \mathcal{N}( {\mu},\Sigma) \quad \quad
N_i  \sim \mathcal{N}(0,{\sigma}^2 I_d )
\end{eqnarray}
We note that while the assumption of a Gaussian distribution does not hold in practice, it facilitates the derivation of the new measure. The justification of the new measure is its superiority over the existing class averaging algorithm, as we demonstrate in Sec. \ref{sec:num}.

The Gaussian assumption on signal and noise (5) and the image formation model (2) imply that $Y_i$ is also Gaussian
\begin{equation}
Y_i \sim \mathcal{N}(A_i \mu, A_i \Sigma A_i^T + \sigma^2 I_d),\quad \text{for } i=1,\ldots,n.
\end{equation} 
%We denote the covariance of $Y_i$ by $K_i$.
%\begin{eqnarray}
%% E(Y_1)=H_1 \mu \\ 
%Cov(Y_i) = A_i \Sigma {A_i}^T + {\sigma}^2 I_n = K_i,\quad \text{for} \quad i=1,\ldots,n 
%%E(Y_2)=H_2 \mu \\
%%Cov(Y_2) = H_2 \Sigma {H_2}^T + {\sigma}^2 I_n = K_2 
%\end{eqnarray}
%Using the Gaussian property, we have the following probability density functions (pdf) for $i=1,\ldots,n$
%\begin{equation}
%\small{
%f_{X_i}(x_i) = P \exp\{-\frac{1}{2}{(x_i-\mu)}^T {\Sigma}^{-1}(x_i-\mu)\},}
%\end{equation}
%%\begin{equation}
%%\small{f_{X_2}(x_2) = P \exp\{-\frac{1}{2}{(x_2-\mu)}^T {\Sigma}^{-1}(x_2-\mu)\} }
%%\end{equation}
%\begin{equation}
%\small{f_{N_i}(z_i) = Q \exp\{-\frac{1}{2}{z_i}^T \frac{1}{\sigma^2}z_i\},}
%\end{equation}
%%\begin{equation}
%%\small{f_N(z_2) = Q \exp\{-\frac{1}{2}{(z_2)}^T \frac{1}{\sigma^2} (z_2)\}, z_2=y_2-H_2 x_2,}
%%\end{equation}
%\begin{equation}
%\small{f_{Y_i}(y_i) = R_i \exp\{-\frac{1}{2}{(y_i-A_i\mu)}^T {K_1}^{-1}(y_i-A_i\mu)\}, }
%\end{equation}
%%\begin{equation}
%%\small{f_{Y_2}(y_2) = R \exp\{-\frac{1}{2}{(y_2-H_2\mu)}^T {K_2}^{-1}(y_2-H_2\mu)\}}
%%\end{equation}
%where $P=\frac{1}{({2\pi})^{\frac{d}{2}}{|\Sigma|}^{\frac{1}{2}}}$, $Q=\frac{1}{({2\pi})^{\frac{d}{2}}\sigma^n}$, and $R_i=\frac{1}{({2\pi})^{\frac{d}{2}} {|K_1|}^{\frac{1}{2}}}$.
The joint distribution of $(X_i, Y_i)$ is given by
\begin{equation}
 \left[\begin{array}{c} X_i \\ Y_i \end{array}\right] = 
\begin{bmatrix} I & 0 \\ A_i & I \end{bmatrix} \times \left[ \begin{array}{c} X_i \\ N_i \end{array} \right]      
\end{equation}
\begin{equation}
 \sim \mathcal{N} \left[\begin{bmatrix} \mu \\ A_i \mu \end{bmatrix}, \begin{bmatrix} \Sigma & \Sigma A_i^T \\ A_i \Sigma & A_i \Sigma A_i^T + \sigma^2 I \end{bmatrix} \right]
\end{equation}
\noindent
The conditional distribution of $X_i$ given $Y_i$ is also Gaussian 
 \begin{equation}
  {X_i|Y_i} = y_i \sim \mathcal{N}(\alpha_i, L_i)
 \end{equation}
%\begin{equation}
%  f_{X_2|Y_2}(x_2|y_2) \sim N(\beta, M)
% \end{equation}
where 
\begin{equation}
\small{
\begin{array}{l}\alpha_i = \mu + \Sigma A_i^T (A_i \Sigma A_i^T + \sigma^2 I)^{-1} (y_i - A_i \mu) \\ L_i = \Sigma - \Sigma A_i^T (A_i \Sigma A_i^T + \sigma^2 I)^{-1} A_i \Sigma .
       \end{array}}  
\end{equation}
So
\begin{equation}
X_i - X_j|Y_i=y_i, Y_j=y_j \sim \mathcal{N}(\alpha_i-\alpha_j, L_i+L_j)
\end{equation}
Let $X_i - X_j = x_{ij}$, and $\alpha_i -\alpha_j=\alpha_{ij}$. Then, for small $\epsilon$, the probability that the $\ell_p$ distance between $X_i$ and $X_j$ is smaller than $\epsilon$ is
\begin{align}
&\Pr(||X_{ij}||_{p} < \epsilon|Y_i=y_i,Y_j=y_j)   \nonumber
%										 =	 %\Pr(||x_{ij}||_{p} < \epsilon|y_i,y_j)\\ %%\nonumber
										=   \frac{1}{(2 \pi)^{\frac{d}{2}} |L_{i}+L_{j}|^ \frac{1}{2}} \times \\ 
										& \int_{B_p(0,\epsilon) }^{}\exp \{ -\frac{1}{2}(x_{ij} - \alpha_{ij})^T(L_i+L_j)^{-1}(x_{ij} - \alpha_{ij}\}dx_{ij} 
\end{align}
\begin{equation}
= \frac{\epsilon^d \text{Vol}(B_p(0,1)) }{(2 \pi)^{\frac{d}{2}} |L_i + L_j|^{\frac{1}{2}}} \exp\{-\frac{1}{2}\alpha_{ij}^T(L_i+L_j)^{-1}\alpha_{ij}\} +  \mathcal{O}(\epsilon^{d+1}) 
\label{eqn:metr}
\end{equation}
where $B_p(0,1)$ is the $\ell_p$ ball of radius $1$ in $\mathbb{R}^d$ centered at the origin (which follows from Taylor expanding the integrand). The probability of $\|X_{ij}\|_p < \epsilon$ given the noisy images $y_i$ and $y_j$  is a measure of the likelihood for the underlying clean images $x_i$ and $x_j$ to originate from the same viewing direction. We define our similarity measure after taking the logarithm on both sides of eqn. \ref{eqn:metr}), dropping out the constants independent of $i$ and $j$, and substituting back $\alpha_{ij}$:
\begin{equation}\label{eq:mah_eqn}
 -\frac{1}{2}\log(|L_i + L_j|) -\frac{1}{2}(\alpha_i - \alpha_j)^T(L_i+L_j)^{-1}(\alpha_i -\alpha_j)
\end{equation}

Notice the resemblance of the second term in eq.~\eqref{eq:mah_eqn} to the classical Mahalanobis distance \cite{mah}. This term takes into account the anisotropic nature of the covariance matrix by appropriately normalizing each dimension when computing the distance between two points. Note that this distance is different for different pairs of points since it depends on $L_i + L_j$, unlike the Euclidean distance and the classical Mahalanobis distance. 
%Upto the first term, the similarity measure %defined here is closely related to the one in \cite{nlica}. 
\vspace{-.15in}
\section{Algorithm for Improved Class Averaging using Mahalanobis Distance}

We propose an improved class averaging algorithm that incorporates the affinity measure~\eqref{eq:mah_eqn}. The quantities $\alpha_i$, $L_i$ are computed for each image and defocus group respectively (in practice $\Sigma$ is replaced by its estimate $\hat{\Sigma}$), using CWF \cite{cwf}. The estimated covariance using CWF is block diagonal in the Fourier Bessel basis. In practice, we use $\alpha_i$, $L_i$ projected onto the subspace spanned by the principal components (for each angular frequency block). We obtain an initial list of $S$ nearest neighbors for each image using the Initial Classification algorithm in \cite{zhao}. Then, for the list of nearest neighbors corresponding to each image, the affinity~\eqref{eq:mah_eqn} is computed and used to pick the closest $K$ nearest neighbors, where $K<S$. 
%
%\begin{algorithm}
%\caption{Improved Class Averaging }
%\label{alg:classav}
%\begin{algorithmic}[1]
%\Require  A list of S nearest neighbor suspects for each image using initial classification \cite{zhao}. 
%%\Procedure{Initial Classification \cite{zhao}}{}
%%\State Image compression and denoising: compute Fourier Bessel steerable basis for images \cite{ffbspca}
%%\State Rotationally invariant features: compute the bispectrum from denoised coefficients in the steerable basis
%%\State Randomized PCA\cite{rokhlin} of high dimensional feature vectors from the bispectrum
%%\State Initial nearest neighbor classification and alignment using brute force or fast randomized approximate nearest neighbor search \cite{fastnn}
%%%\State (Optional) Improve nearest neighbor classification using Vector Diffusion Maps (VDM) \cite{vdm}
%%\State Obtain a list of $S$ nearest neighbor suspects for each image
%%\EndProcedure
%\Procedure{Classification using Mahalonobis Distance}{}
%\State Compute the quantities $\alpha_i, L_i$ using estimates from Covariance Wiener Filtering (CWF) \cite{cwf}
%\State For each image and its $S$ aligned nearest neighbors, compute the Mahalanobis distance between the image and neighbors
%\State Rank $S$ suspects according to the Mahalanobis distance, and choose the top $K$ as nearest neighbors
%\EndProcedure
%\Procedure{(Optional) Improve nearest neighbor classification using Vector Diffusion Maps (VDM) \cite{vdm}}{}
%\EndProcedure
%\end{algorithmic}
%\end{algorithm}

\begin{table}[]
\centering
\begin{tabular}{|c|c|c|c|}
       \hline       
SNR   & This work        & \cite{zhao} (VDM)        & \cite{zhao} (No VDM)                  \\ \hline
1/40 &  58373         & 56896          & 49560       \\
1/60  & 34965          & 32113          & 29219       \\
1/100 & 17262      & 14431           & 13706         \\
\hline
\end{tabular}
\caption{Number of nearest neighbors with correlation $>0.9$, 
using 10,000 images, $K=10$ and $S=50$.
\label{table:tab}}
\end{table}
%\begin{figure}[!htbp]
%\begin{center}
%\includegraphics[width=.9 \columnwidth]{classavg_fig3.pdf}
%\end{center}
%\vspace{-.15in}
%\caption{Results of class averaging of a synthetic dataset of $10000$ projection images of size $65 \times 65$, affected by CTF and at an SNR$=1/40$. We show class averages with the improved algorithm using the anisotropic affinity, for $S=50$, and $K=10,20,30$.}
%\vspace{-.25in}
%\label{fig:classavg_k}
%\end{figure}

\section{Numerical experiments}
\label{sec:num}

\begin{figure}
\begin{center}
\begin{tabular}{cc}
\includegraphics[width=.49\columnwidth]{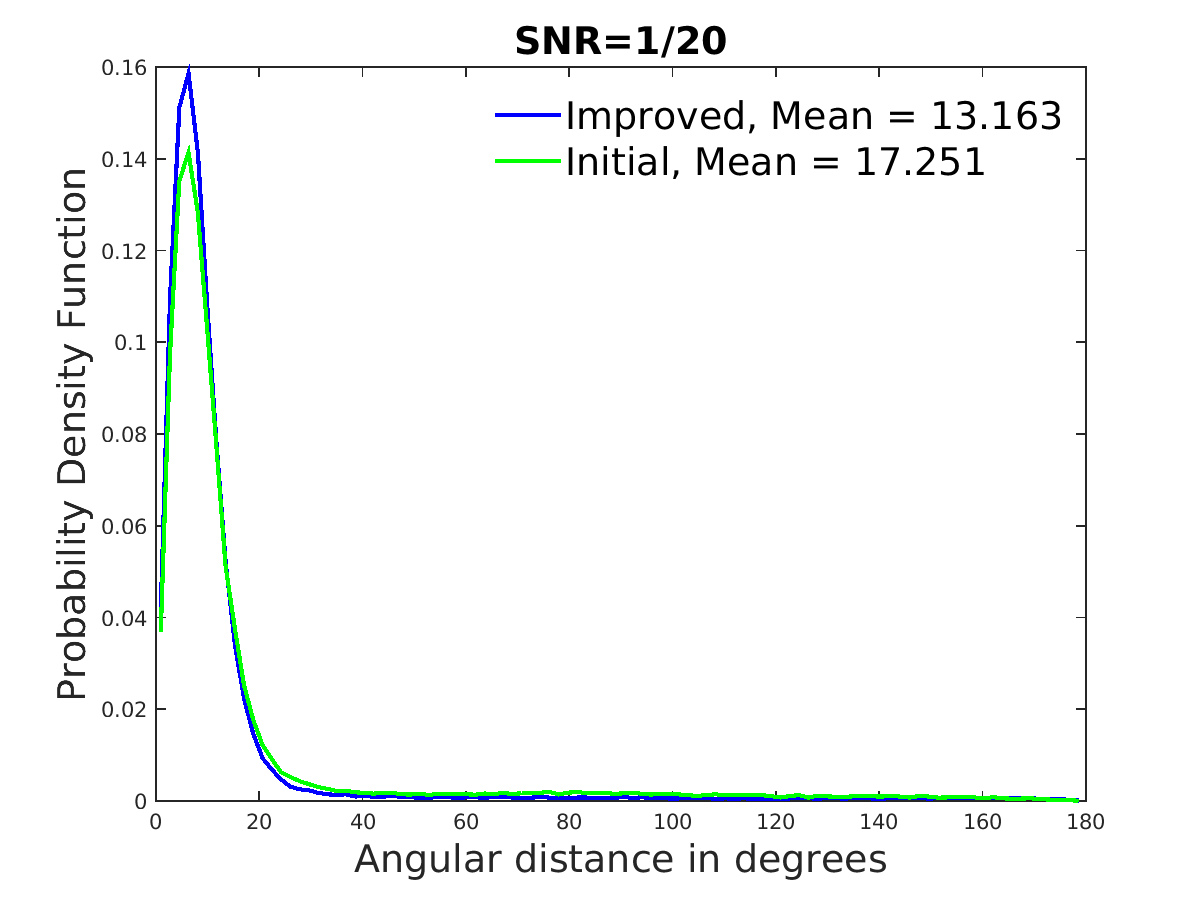} & \includegraphics[width=.49\columnwidth]{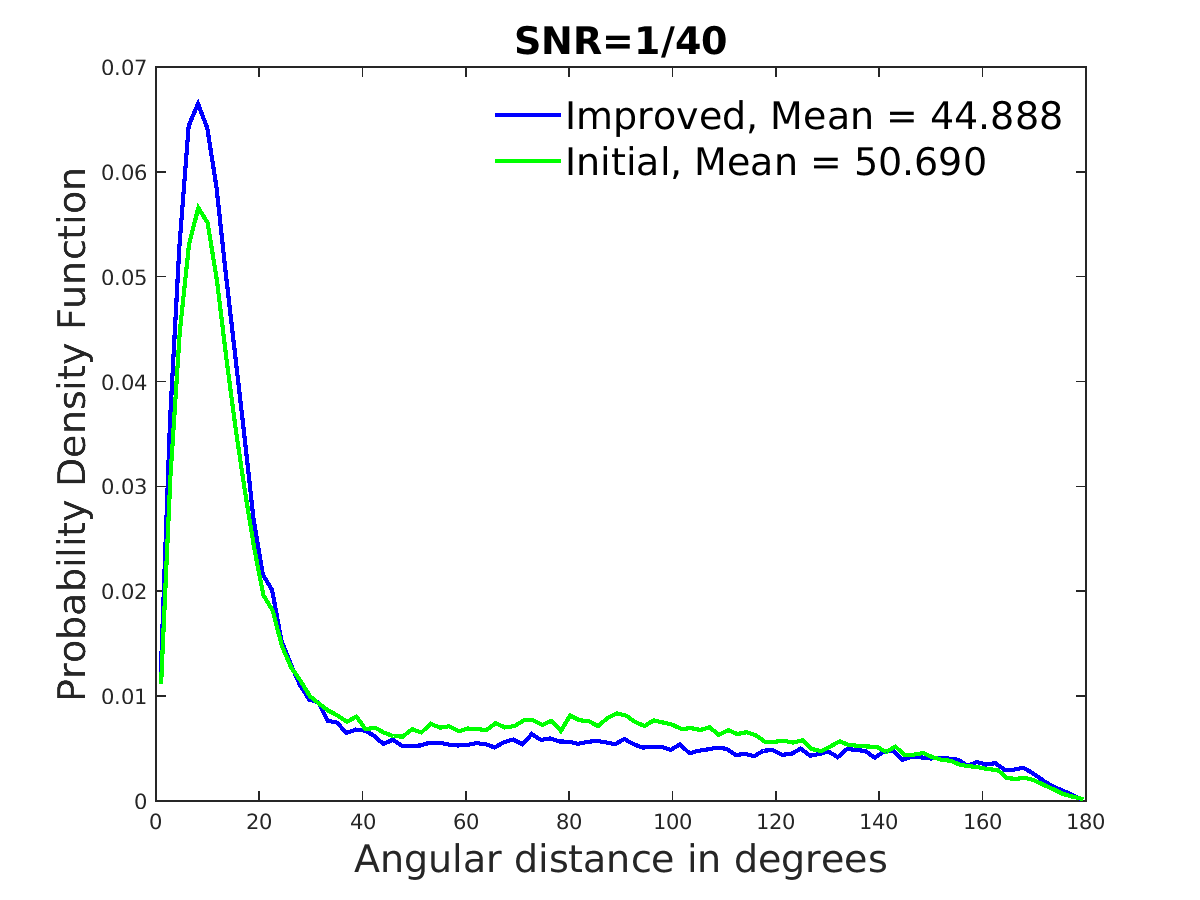} \\
\includegraphics[width=.49\columnwidth]{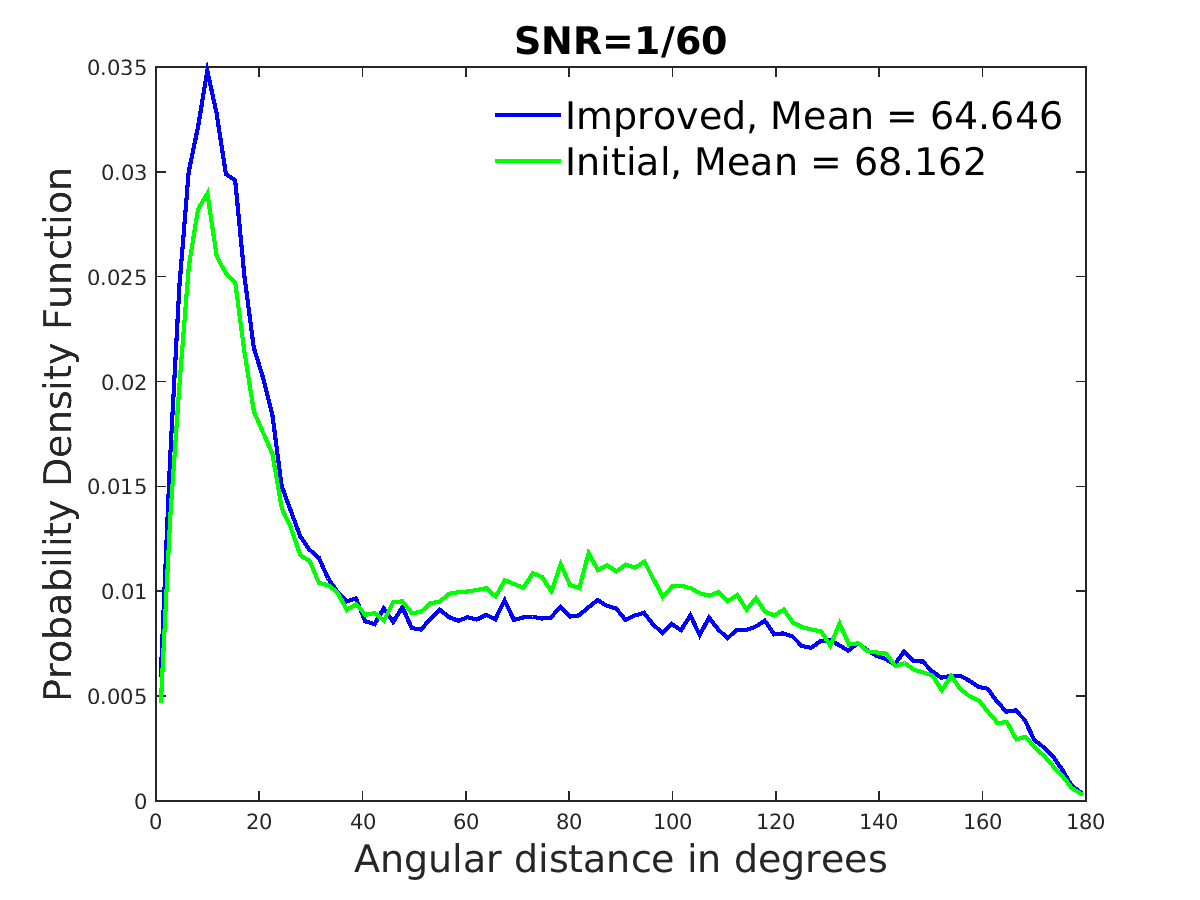} & \includegraphics[width=.49\columnwidth]{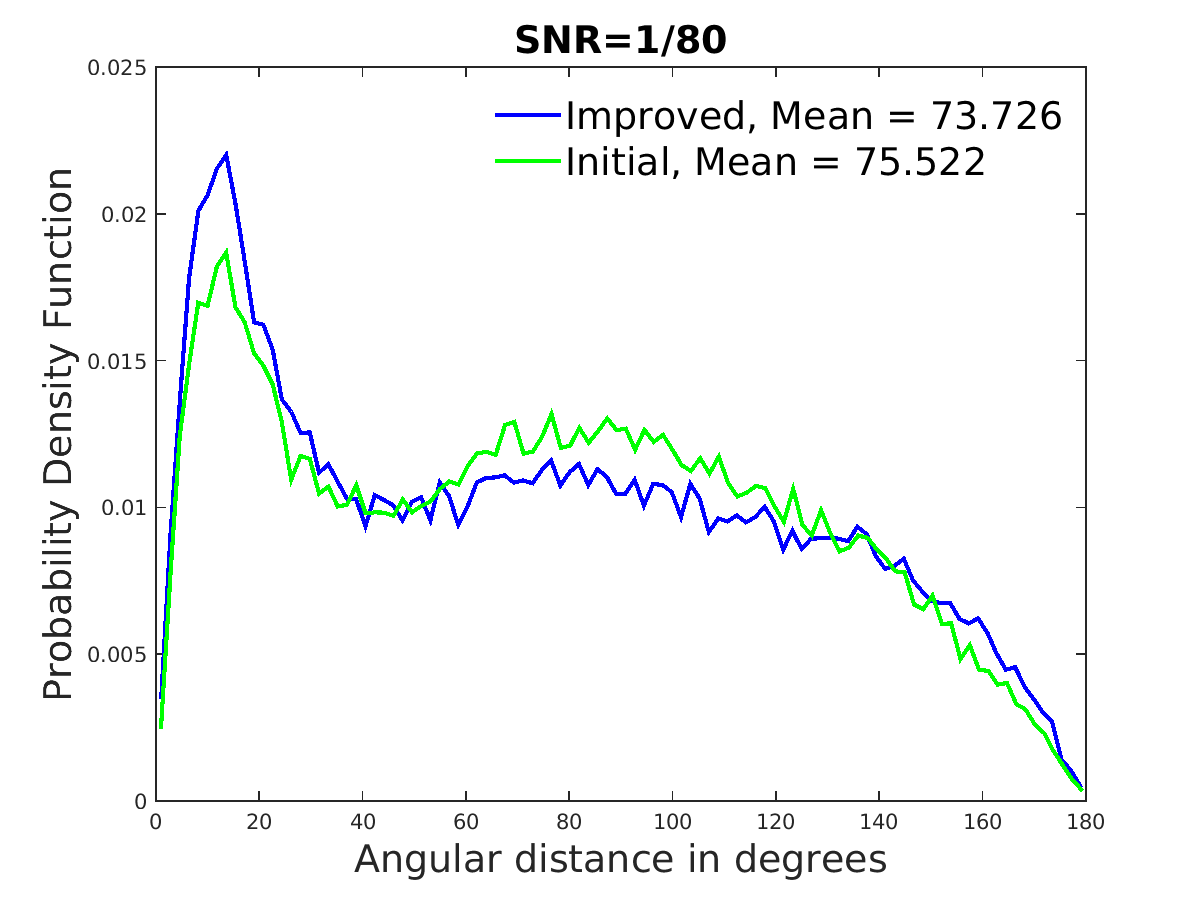}
\end{tabular}
\end{center}
\vspace{-.15in}
\caption{The estimated probability density function of the angular distance (in degrees) between images classified into the same class by 1) Initial Classification and 2) Improved Classification using the anisotropic affinity at different SNR's.}
\vspace{-.15in}
\label{fig:hist}
\end{figure}

\begin{figure}[!htbp]
\begin{center}
\begin{tabular}{c}
\includegraphics[width=.9 \columnwidth]{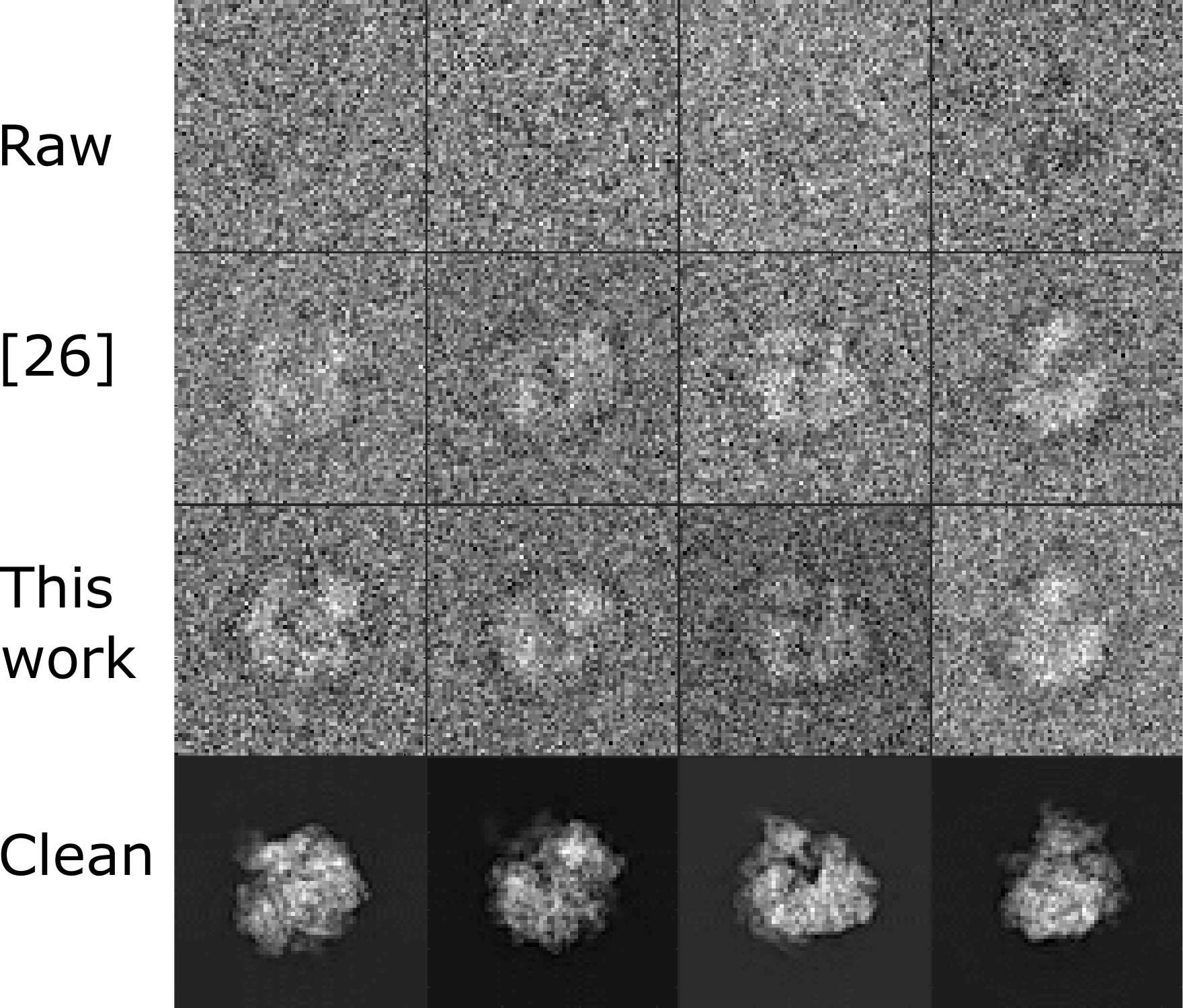}\\
(a)\\
\includegraphics[width=.9 \columnwidth]{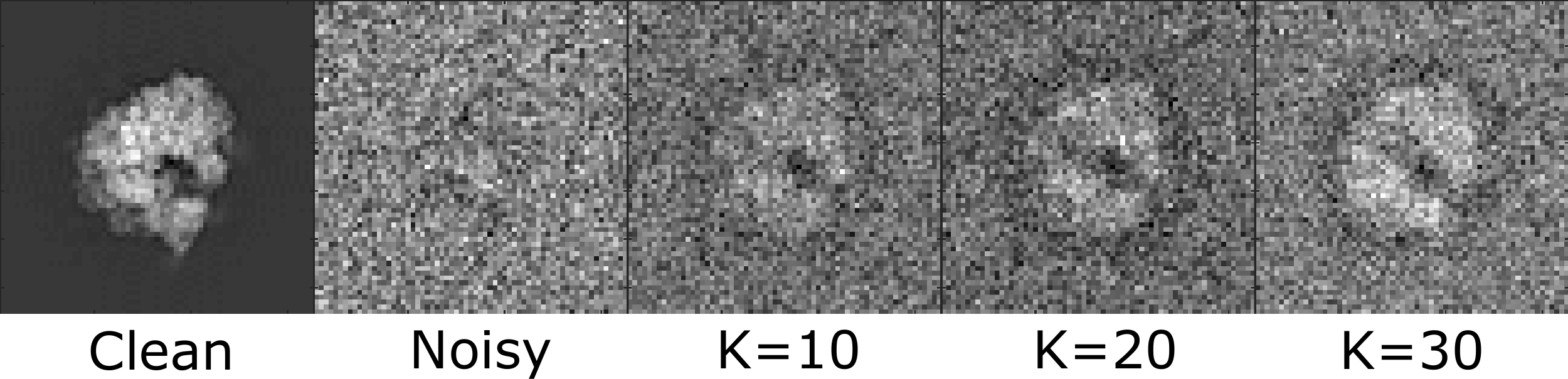}\\
(b)
\end{tabular}
\end{center}
\vspace{-.15in}
\caption{Results of class averaging of a synthetic dataset of $10000$ projection images of size $65 \times 65$, affected by CTF and SNR$\ =1/40$. (a) We show class averages with Initial Classification in the second row, and with the improved algorithm using the anisotropic affinity in the third row. We use $K=10$ and $S=50$. (b) Class averages for one image in the dataset with the improved algorithm using the anisotropic affinity, for $S=50$, and using $K=10,\ 20,\ 30$.}
\vspace{-.25in}
\label{fig:classavg}
\end{figure}

We test the improved class averaging algorithm on a synthetic dataset that consists of projection images generated from the volume of P. falciparum 80S ribosome bound to E-tRNA, available on the Electron Microscopy Data Bank (EMDB) as EMDB 6454. The algorithm was implemented in the UNIX environment, on a machine with total RAM of 1.5 TB, running at 2.3 GHz, and with 60 cores. 
For the results described here, we used $10,000$ projection images of size $65 \times 65$ that were affected by the various CTF's and additive white Gaussian noise at various noise levels, in particular, we show here results for 4 values of the SNR. The images were divided into $20$ defocus groups. Initial classification was first used to select $S=50$ nearest neighbors for each image. After rotationally aligning the suspected neighbors, the Mahalanobis distance was computed between each image and the $50$ aligned suspects. We then pick the closest $K=10$ neighbors for each image (in practice, the choice of $K$ depends on the SNR and the number of images). For comparison, we compute $10$ nearest neighbors for each image using only Initial Classification (with or without using the optional VDM step). Table \ref{table:tab} shows the number of pairs of nearest neighbor images detected with each method at various SNR's, that have correlation $>0.9$ between the original clean images, indicating that they are indeed neighbors. We note an improvement in the number of true nearest neighbors detected by the improved classification algorithm using the Mahalanobis distance. Figure \ref{fig:hist} shows the estimated probability density function of the angular distance between nearest neighbor images, using 1) Initial Classification only 2) Improved classification using the Mahalanobis distance by repeating this experiment at four different SNR's. Figure \ref{fig:classavg}a shows the results of Initial Classification and the improved class averaging algorithm on this synthetic dataset. We compare the quality of the class averages from \cite{zhao} and this paper with $K=10$. Averaging over a large number of nearest neighbors reduces the noise variance. However, it also blurs the underlying clean signal, since the neighbors are not exactly from the same viewing direction. Therefore, it is crucial to correctly identify only the top few nearest neighbors and average them in order to sufficiently reduce the noise without  blurring the features too much. We see in Figure \ref{fig:classavg}b that noise reduces in the class averages when $K$ increases. The procedure in \cite{zhao} took $168$ seconds, while the improved classification using the anisotropic affinity took $860$ seconds for the experiment described here. 

\vspace{-.15in}
\section{Discussion}
We introduced a new similarity measure to compare CTF-affected cryo-EM images belonging to different defocus groups.  The anisotropic affinity derived in this paper is similar to the one that appears in \cite{nlica, intgeom} but also includes an additional logarithmic term. We provided a new probabilistic interpretation for this anisotropic affinity. The affinity can also be used as a similarity measure for any manifold learning procedure \cite{intgeom, nlica} such as diffusion maps \cite{vdm, difmap}, with or without missing data, and extended to other imaging modalities where images are affected by different point spread functions or blurring kernels.
\bibliographystyle{abbrv}
\bibliography{cwf_isbi_ref}

\begin{thebibliography}{10}

\bibitem{cwf}
T.~Bhamre, T.~Zhang, and A.~Singer.
\newblock Denoising and covariance estimation of single particle cryo-{EM}
  images.
\newblock {\em Journal of Structural Biology}, 195(1):72 -- 81, 2016.

\bibitem{difmap}
R.~Coifman and S.~Lafon.
\newblock Diffusion maps.
\newblock {\em Applied and Computational Harmonic Analysis}, 21(1):5 -- 30,
  2006.

\bibitem{Dube1993}
P.~Dube, P.~Tavares, R.~Lurz, and M.~van Heel.
\newblock Bacteriophage {SPP}1 portal protein: a {DNA} pump with 13-fold
  symmetry.
\newblock {\em EMBO J.}, 15:1303--1309, 1993.

\bibitem{frankbook}
J.~Frank.
\newblock Electron microscopy of macromolecular assemblies.
\newblock In J.~Frank, editor, {\em Three-Dimensional Electron Microscopy of
  Macromolecular Assemblies}, pages 12 -- 53. Academic Press, Burlington, 1996.

\bibitem{Frank1}
J.~Frank.
\newblock {\em Three-Dimensional Electron Microscopy of Macromolecular
  Assemblies : Visualization of Biological Molecules in Their Native State:
  Visualization of Biological Molecules in Their Native State}.
\newblock Oxford University Press, USA, 2006.

\bibitem{gpurelion}
D.~Kimanius, B.~O. Forsberg, S.~Scheres, and E.~Lindahl.
\newblock Accelerated cryo-{EM} structure determination with parallelisation
  using {GPU}s in {RELION}-2.
\newblock {\em bioRxiv}, 2016.

\bibitem{Kuhlbrandt1443}
W.~K{\"u}hlbrandt.
\newblock The resolution revolution.
\newblock {\em Science}, 343(6178):1443--1444, 2014.

\bibitem{mah}
P.~C. Mahalanobis.
\newblock {On the generalised distance in statistics}.
\newblock In {\em Proceedings National Institute of Science, India}, volume~2,
  pages 49--55, Apr. 1936.

\bibitem{xmipp}
R.~Marabini, I.~M. Masegosa, M.~S. Mart\'in, S.~Marco, J.~Fernández, L.~de~la
  Fraga, C.~Vaquerizo, and J.~Carazo.
\newblock Xmipp: An image processing package for electron microscopy.
\newblock {\em Journal of Structural Biology}, 116(1):237 -- 240, 1996.

\bibitem{chirikjian1}
W.~Park and G.~Chirikjian.
\newblock An assembly automation approach to alignment of noncircular
  projections in electron microscopy.
\newblock {\em IEEE Transactions on Automation Science and Engineering},
  11(3):668--679, 1 2014.

\bibitem{chirikjian2}
W.~Park, C.~R. Midgett, D.~R. Madden, and G.~S. Chirikjian.
\newblock A stochastic kinematic model of class averaging in single-particle
  electron microscopy.
\newblock {\em I. J. Robotics Res.}, 30:730--754, 2011.

\bibitem{Penczek1992}
P.~A. Penczek, M.~Radermacher, and J.~Frank.
\newblock Three-dimensional reconstruction of single particles embedded in ice.
\newblock {\em Ultramicroscopy}, 40:33--53, 1992.

\bibitem{Penczek1996}
P.~A. Penczek, J.~Zhu, and J.~Frank.
\newblock A common-lines based method for determining orientations for {$N>3$}
  particle projections simultaneously.
\newblock {\em Ultramicroscopy}, 63(3-4):205--218, 1996.

\bibitem{rokhlin}
V.~Rokhlin, A.~Szlam, and M.~Tygert.
\newblock A randomized algorithm for principal component analysis.
\newblock {\em SIAM Journal on Matrix Analysis and Applications},
  31(3):1100--1124, 2010.

\bibitem{vanHeel1990a}
M.~Schatz and M.~van Heel.
\newblock Invariant classification of molecular views in electron micrographs.
\newblock {\em Ultramicroscopy}, 32:255--264, 1990.

\bibitem{relion}
S.~H.~W. Scheres.
\newblock {RELION}: Implementation of a bayesian approach to cryo-em structure
  determination.
\newblock {\em Journal of Structural Biology}, 180(3):519 -- 530, 2012.

\bibitem{spider}
T.~R. Shaikh, H.~Gao, W.~T. Baxter, F.~J. A., N.~Boisset, A.~Leith, and
  J.~Frank.
\newblock {SPIDER} image processing for single-particle reconstruction of
  biological macromolecules from electron micrographs.
\newblock {\em Nature Protocols}, 3(12):1941--1974, 2008.

\bibitem{nlica}
A.~Singer and R.~R. Coifman.
\newblock Non-linear independent component analysis with diffusion maps.
\newblock {\em Applied and Computational Harmonic Analysis}, 25(2):226 -- 239,
  2008.

\bibitem{vdm}
A.~Singer and H.-T. Wu.
\newblock Vector diffusion maps and the connection {L}aplacian.
\newblock {\em Communications on Pure and Applied Mathematics},
  65(8):1067--1144, 2012.

\bibitem{intgeom}
R.~Talmon and R.~R. Coifman.
\newblock Empirical intrinsic geometry for nonlinear modeling and time series
  filtering.
\newblock {\em Proceedings of the National Academy of Sciences},
  110(31):12535--12540, 2013.

\bibitem{eman2}
G.~Tang, L.~Peng, P.~R. Baldwin, D.~S. Mann, W.~Jiang, I.~Rees, and S.~J.
  Ludtke.
\newblock {EMAN2}: An extensible image processing suite for electron
  microscopy.
\newblock {\em Journal of Structural Biology}, 157:38 -- 46, 2007.
\newblock Software tools for macromolecular microscopy.

\bibitem{vanHeel1981}
M.~{van Heel} and J.~Frank.
\newblock Use of multivariate statistics in analysing the images of biological
  macromolecules.
\newblock {\em Ultramicroscopy}, 6(2):187--194, 1981.

\bibitem{imagic}
M.~{van Heel}, G.~Harauz, E.~V. Orlova, R.~Schmidt, and M.~Schatz.
\newblock A new generation of the {IMAGIC} image processing system.
\newblock {\em Journal of Structural Biology}, 116(1):17 -- 24, 1996.

\bibitem{mahclust1}
S.~Xiang, F.~Nie, and C.~Zhang.
\newblock Learning a mahalanobis distance metric for data clustering and
  classification.
\newblock {\em Pattern Recognition}, 41(12):3600 -- 3612, 2008.

\bibitem{mahclust2}
X.~Zhao, Y.~Li, and Q.~Zhao.
\newblock Mahalanobis distance based on fuzzy clustering algorithm for image
  segmentation.
\newblock {\em Digit. Signal Process.}, 43(C):8--16, Aug. 2015.

\bibitem{zhao}
Z.~Zhao and A.~Singer.
\newblock Rotationally invariant image representation for viewing direction
  classification in cryo-{EM}.
\newblock {\em Journal of Structural Biology}, 186(1):153 -- 166, 2014.

\end{thebibliography}

\end{document}